\documentclass[11pt]{article}
\input epsf.tex
\newdimen\SaveWidth \SaveWidth=\textwidth
\newdimen\SaveHeight \SaveHeight=\textheight
\advance\SaveWidth by -\textwidth
\advance\SaveHeight by -\textheight

\divide\SaveWidth by 2
\divide\SaveHeight by 2
\advance\hoffset by \SaveWidth
\advance\voffset by \SaveHeight

\def\ie{\it i.e.}
\def\eg{\it e.g.}

\let\badcite=\cite
\def\cite{~\badcite}

\def\slashchar#1{\setbox0=\hbox{$#1$}           
   \dimen0=\wd0                                 
   \setbox1=\hbox{/} \dimen1=\wd1               
   \ifdim\dimen0>\dimen1                        
      \rlap{\hbox to \dimen0{\hfil/\hfil}}      
      #1                                        
   \else                                        
      \rlap{\hbox to \dimen1{\hfil$#1$\hfil}}   
      /                                         
   \fi} 
    \def\slashword#1{\setbox0=\hbox{$#1$}        
  \dimen0=\wd0                                   
   \setbox1=\hbox{/} \dimen1=\wd1                
   \ifdim\dimen0>\dimen1                         
      \rlap{\hbox to \dimen0{\hfil\bf---\hfil}} %
      #1                                         %
   \else                                         
      \rlap{\hbox to \dimen1{\hfil$#1$\hfil}}    
      /                                          
    \fi}                                         %

\catcode`@=11
\newdimen\vbigd@men                             

\def\vbig#1#2{{\vbigd@men=#2\divide\vbigd@men by 2%
   \hbox{$\left#1\vbox to \vbigd@men{}\right.\n@space$}}}
\catcode`@=12

\catcode`@=11
\def\citenum#1{\csname b@#1\endcsname}
\catcode`@=12

\def\dofig#1#2{\centerline{\epsfxsize=#1\epsfbox{#2}}}
\begin{document}
\begin{titlepage}
\rightline{TUHEP-TH-03144}

\bigskip\bigskip

\begin{center}{\Large\bf\boldmath
Nonsymmetric Gravity and Noncommutative Signals\footnotemark \\}
\end{center}
\footnotetext{ This work was supported by the Department of Physics at Tsinghua University.
}
\bigskip
\centerline{\bf N. Kersting$^{a}$,
		 and Y.L. Ma$^{a}$}
\centerline{{\it$^{a}$ Theoretical Physics Group, Department of Physics
		       Tsinghua University, Beijing P.R.C. 100084
		       }}
\bigskip

\begin{abstract}

	Models in which the space-time metric $g_{\mu\nu}$ is not symmetric, 
$\ie$  $g_{\mu\nu} \ne g_{\nu\mu}$ may make predictions in
 scattering experiments, for example in a future $e^+e^-$ linear collider,
 similar to those from 
noncommutative field theory. We compute the differential cross sections
for pair annihilation, Bhabha and M$\o$ller scattering and
find that both nonsymmetric gravity theory(NGT) and
noncommutative field theory  predict a similar
dependence of the
differential cross section on the azimuthal angle in agreement with
all known data, however in NGT Lorentz violation need
not be as severe. Astrophysical and cosmological tests may prove
very  useful in
distinguishing these two theories.

\bigskip        

\end{abstract}

\newpage
\pagestyle{empty}

\end{titlepage}

\section{Introduction}
\label{sec:intro}

If Nature violates isotropy of space or
 commutivity among coordinate displacements, the basic assumptions
at the root of most of physics come into question and it becomes
 imperative to parametrize
this violation, devising
tests to put limits on the parameters.
Two such parameterizations in this direction involve the theories of 
noncommuting space coordinates\cite{Hinchliffe:2002km} and nonsymmetric
 gravity\cite{Moffat:1995fc}. Particle physics experiments at
 future hadronic or linear colliders  may
 provide good testing grounds for these theories.

In nonsymmetric gravity theory (NGT) the metric of  space-time is taken to be
nonsymmetric, $\ie$  $g_{\mu\nu} \ne g_{\nu\mu}$. In particular, we may write 
\begin{equation}
\label{g-components}
g_{\mu\nu} = g_{(\mu\nu)} + g_{[\mu\nu]}
\end{equation}
decomposing $g$ into its symmetric and antisymmetric pieces.
The contravariant tensor $g^{\mu\nu}$ is defined as usual:
\begin{equation}
g^{\mu\nu}g_{\mu\rho} = \delta^\nu_\rho 
\end{equation}
One can now go on to define a Lagrangian density as in general relativity,
${\cal L} = \sqrt{-g} R$, where $g\equiv det(g_{\mu\nu})$ and
$R$ is the Ricci scalar,  and derive field equations for $g_{(\mu\nu)}$ 
and $g_{[\mu\nu]}$. More details on this reformulation of general
relativiy can be found in \cite{Moffat:1995fc}.
 There has been extensive work
analyzing the effects of $g_{[\mu\nu]}$ for black hole solutions of the
field equations, galaxy dynamics, stellar stability, and other phenomena
of cosmological and astrophysical relevance 
\cite{Moffat:1997cc,Moffat:1995pi,Moffat:1996dq} where
$g_{(\mu\nu)}$ and $g_{[\mu\nu]}$  may be of comparable size.

In the context of particle physics however, we may start with the
assumption that the curvature of
space in the region of interest is small:
\begin{equation}
\label{g-defn}
g_{\mu\nu} \approx \eta_{\mu\nu} +  h_{(\mu\nu)} + a_{[\mu\nu]}
\end{equation}
where $\eta$ is the usual Minkowski metric and both\footnotemark~ $h$ and $a$
 \footnotetext{Note that $a_{\mu\nu}$ cannot
be absorbed into $\eta_{\mu\nu}$ or $h$ by a redefinition of coordinates}
satisfy
$a_{\mu\nu}, h_{\mu\nu} \ll 1 ~\forall~ \mu,\nu$.
We further assume that these fields' dynamics are negligable in the
region of interest and we may treat them as background fields.
The effects of the symmetric tensor $h$ on particle physics in this
limit has been studied elsewhere 
(see for example \cite{h-study,Gusev:1998rp,DiPiazza:2003zp}).
 We would like to
focus our attention here on the effects
of the antisymmetric piece $a$.

We therefore take  $h=0$ and
  $a_{\mu\nu} = {\cal O}(\epsilon) \ll 1 ~\forall~ \mu,\nu$.
 To simplify computations, we take the form of $a$ to be
\begin{equation}
\label{a-form}
a_{\mu\nu} =  
	  \left( \begin{array}{cccc} 
	0 & \epsilon  & \epsilon & \epsilon\\
	-\epsilon & 0  & \epsilon & \epsilon\\
	-\epsilon & -\epsilon  & 0 & \epsilon\\
	-\epsilon & -\epsilon  & -\epsilon  & 0\\
 \end{array} \right)
\end{equation}
In this scenario Eqn(\ref{g-defn}) states that the NGT is a perturbation of the
ordinary flat-space theory in the small parameter $\epsilon$.
This parameter
may depend on space-time, as one would expect from the metric theory
of General Relativity(GR), and from this point on we keep this dependance
implicit in all equations: $\epsilon \equiv \epsilon(x)$, with the
understanding that odd powers of $\epsilon$ appearing in physical
quantities may average to zero over sufficiently large\footnotemark 
regions of space or time.
\footnotetext{The parameter $\epsilon$ might for example vary 
appreciably only between the atomic and sub-micron
scales, justifying its approximate constancy at collider energies
in excess of several $GeV$}

The theory of noncommutating space-time coordinates
has already received much attention
in the literature(see \cite{Hinchliffe:2002km} for an extensive review). Here
we briefly recall its key features.

Noncommutative space-time is a deformation of ordinary space-time in which the
 space-time coordinates $x_\mu$, representable by Hermitian operators $\widehat{x}_\mu$, do not
commute:
\begin{equation}
\label{nceqn}
[\widehat{x}_\mu,\widehat{x}_\nu]=i \theta_{\mu \nu}
\end{equation}
Here $\theta_{\mu \nu}$ is the deformation parameter: ordinary space-time is obtained in the
 $\theta_{\mu \nu} \to 0$ limit.  By convention
it is a real tensor antisymmetric
 under $\mu \leftrightarrow \nu$.
To obtain a noncommuting version of a particular field theory,
 one need only replace ordinary
products between fields with the so-called ``star product'' defined
as :
\begin{equation}
\label{star}
(f \star g)(x) \equiv e^{i \theta_{\mu \nu} \partial_{\mu}^{y} \partial_{\nu}^{z}}
		f(y) g(z) \mid_{y=z=x}
\end{equation}
In particular, one can transform the Standard Model into
 a noncommutative Standard Model (ncSM).
 Noncommuting coordinates are
found to follow naturally in the context of string theory, where
$\theta_{\mu \nu}$ is related to a background electric field. 
The direction of this field explicitly breaks Lorentz invariance,
strongly constraining the size of $\theta_{\mu \nu}$.
 Phenomenological constraints
on the
ncSM, ranging from Hydrogen spectra, $e^+e^-$ scattering, and various
CP-violating quantities\cite{Hinchliffe:2002km}
 imply that the dimensionful parameters
$\theta_{\mu\nu}$ should not exceed $1~(TeV)^{-2}$\footnotemark.
\footnotetext{In some considerations in nuclear physics
 this limit can be pushed many orders of magnitude stronger,
 however this assumes that
$\theta_{\mu\nu}$ is constant over solar-system 
scales\cite{Mocioiu:2001nz}}

In this paper we demonstrate that some signals of NGT at a high
center-of-mass collider bear resemblance
to those of the ncSM. 
We present some calculations in NGT of simple QED processes
 in Section \ref{sec:predict}, showing that differential scattering
cross sections have an oscillatory dependance on the azimuthal
angle $\varphi$ similar to that in the ncSM. We also explicitly
show agreement of NGT with known data from various electron scattering
experiments.
In Section \ref{sec:constraints} we collect some results in a purely
general relativistic formulation, investigating classical constraints
on the theory from Newton's Laws and cosmological considerations. 
Section \ref{sec:concl} contains our conclusions.

\section{Predictions in Simple Processes}
\label{sec:predict}
\subsection{Pair Annihilation: $e^+e^- \to \gamma\gamma$}
\begin{figure}[t]
\dofig{3.50in}{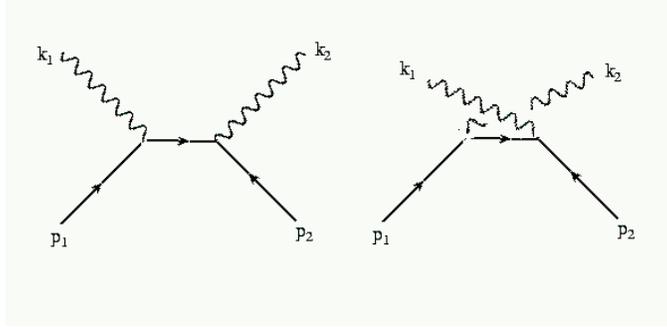}
\caption{ \it The two lowest order diagrams for pair 
annihilation.
\label{ann-fig} }
\end{figure}

Pair annihilation occurs to lowest order in $\alpha_{em}$ through
the two tree-level diagrams shown in Figure \ref{ann-fig}.
As in the SM, the spin-averaged squared amplitude is written
\begin{eqnarray}
\label{pair-cross}
{1\over4}\sum_{spins}|{\cal
M}|^2&=&{e^4\over4}g_{\mu\rho}g_{\nu\sigma}\cdot
tr\bigg\{(p\hspace{-0.17cm}\slash_1+m)\bigg[\frac{\gamma^\mu
k\hspace{-0.17cm}\slash_1\gamma^\nu-2\gamma^\mu p_1^\nu}{2p_1\cdot
k_1}+\frac{\gamma^\nu
k\hspace{-0.17cm}\slash_2\gamma^\mu-2\gamma^\nu
p_1^\mu}{2p_1\cdot k_2}\bigg]\nonumber\\
&&\cdot(p\hspace{-0.17cm}\slash_1+m)\bigg[\frac{\gamma^\sigma
k\hspace{-0.17cm}\slash_1\gamma^\rho-2\gamma^\rho p_1^\sigma}{2p_1\cdot
k_1}+\frac{\gamma^\rho
k\hspace{-0.17cm}\slash_2\gamma^\sigma-2\gamma^\sigma
p_1^\rho}{2p_1\cdot k_2}\bigg]\bigg\}\nonumber\\
\end{eqnarray}
with $m$ the electron mass. Recall the metric tensors are however not as in the SM, but
rather $g_{\mu\nu} = \eta_{\mu\nu} + a_{\mu\nu}$ with
$a_{\mu\nu}$ defined as in Eqn \ref{a-form}. We defer the 
full calculation to the Appendix and simply state our result
for the  differential cross section:
\begin{equation}
\label{pair-cross}
\frac{d\sigma}{d\Omega}_{pSM} + \frac{d\sigma}{d\Omega}_{p1}
+ \frac{d\sigma}{d\Omega}_{p2} 
\end{equation}
where 
\begin{eqnarray}
\frac{d\sigma}{d\Omega}_{pSM} & = & \frac{\alpha^2}{ s}
\bigg[\frac{1 + \cos^2\theta}{\sin^2\theta}\bigg]\nonumber\\
&&\nonumber\\
\frac{d\sigma}{d\Omega}_{p1} &= &\epsilon\frac{\alpha^2}{s}
\bigg[\frac{2\cos\theta}{\sin^2\theta}+\frac{(\sin\varphi+\cos\varphi)}{\sin\theta}\bigg]\nonumber\\
&&\nonumber\\
\frac{d\sigma}{d\Omega}_{p2} &=& 2\epsilon^2 \frac{\alpha^2}{s}
\bigg[\frac{\cos^2\theta-3}{\sin^2\theta}+\frac{(1+\cos\theta)\sin2\theta}{\sin^4\theta}(\sin\varphi-\cos\varphi)-\frac{\sin2\theta}{\sin^2\theta}\cos\varphi\bigg] \nonumber\\
\end{eqnarray}
Here we are working in the high-energy limit ($ m_e \approx 0$)
and the usual angles $\theta$ and $\varphi$ parameterize photon direction 
in the center of mass frame. We have written the ${\cal O}(\epsilon)$
and ${\cal O}(\epsilon^2)$ contributions separately because we will later see
in Section \ref{sec:constraints} that constraints
 on Lorentz violation strongly favor the scenario
where  $\epsilon$ (but not necessarily  $\epsilon^2$) average to 
zero over small distances\footnotemark.
\footnotetext{We record these terms in the cross section in the event that
one wishes to work in a strongly Lorentz-violating theory.}
In this case the prediction of NGT for
the differential cross section is
 $\frac{d\sigma}{d\Omega}_{pSM} + \frac{d\sigma}{d\Omega}_{p2}$.
 
\begin{figure}[t]
\dofig{3.50in}{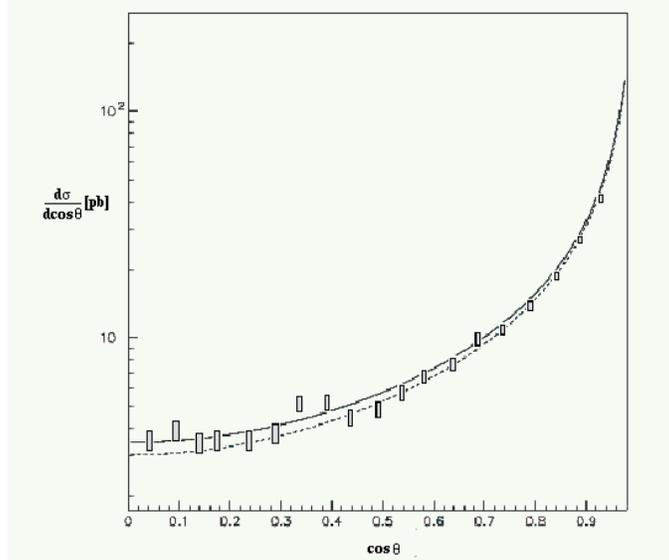}
\caption{\it The OPAL data for $e^+e^- \to \gamma \gamma$ 
\cite{Abbiendi:2003wv} 
shown against the predictions in the SM (solid line) and the
NGT (dashed line) in the case where odd powers of $\epsilon$ average
to zero. Here $\epsilon \approx 0.14$
\label{opal-t-fig} }
\end{figure}

\begin{figure}[t]
\dofig{3.50in}{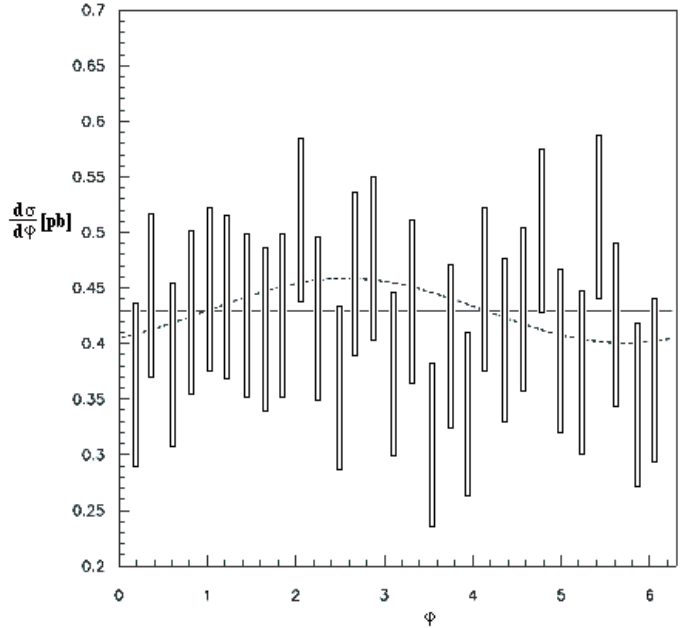}
\caption{\it Differential cross section in the azimuthal angle from 
the OPAL data 
shown against the predictions in the SM (solid line) and the
NGT (dashed line) for $\epsilon \approx 0.14$. An angular
 cut of $cos\theta < 0.6$ was applied.
\label{opal-p-fig} }
\end{figure}

Recently the OPAL Collaboration\cite{Abbiendi:2003wv}
 has analyzed pair annihilation
data at a center of mass energy near $200~GeV$. In Figure
 \ref{opal-t-fig} we show their data
for the differential cross section $d\sigma /d \cos\theta$ and our
prediction $\frac{d\sigma}{d\Omega}_{pSM} + \frac{d\sigma}{d\Omega}_{p2}$.
In this case, $\epsilon$ may be as large as $\approx 0.14$
without deviating more than $\sim~1~\sigma$ with the data.
Using this value of $\epsilon$ we predict the azimuthal 
distribution in Figure \ref{opal-p-fig}, again showing the OPAL data 
alongside the prediction from the SM (in this case a flat line).
As one can see from the figure, the data are consistent with
 both the SM and NGT for this value of $\epsilon$.

The prediction from  noncommutative models similarly consists of 
a negative correction to $d\sigma/d\cos\theta$ and an
oscillatory $d\sigma/d\varphi$ \cite{Hewett:2000zp}. 
 In this case the OPAL data are 
consistent with $\theta_{\mu\nu}< (141~ GeV)^{-2}$, to which we
refer the reader to the original OPAL report for the details.

\subsection{Bhabha Scattering: $e^+e^- \to e^+e^-$}
As seen in the previous section, the $d\sigma/d\cos\theta$
distribution for pair annihiliation served to constrain 
$\epsilon$ more severely than the azimuthal distribution
$d\sigma/d\varphi$. Our aim in this and the following section
is to see whether other simple scattering processes can add
to this constraint.
 
\begin{figure}[t]
\dofig{3.50in}{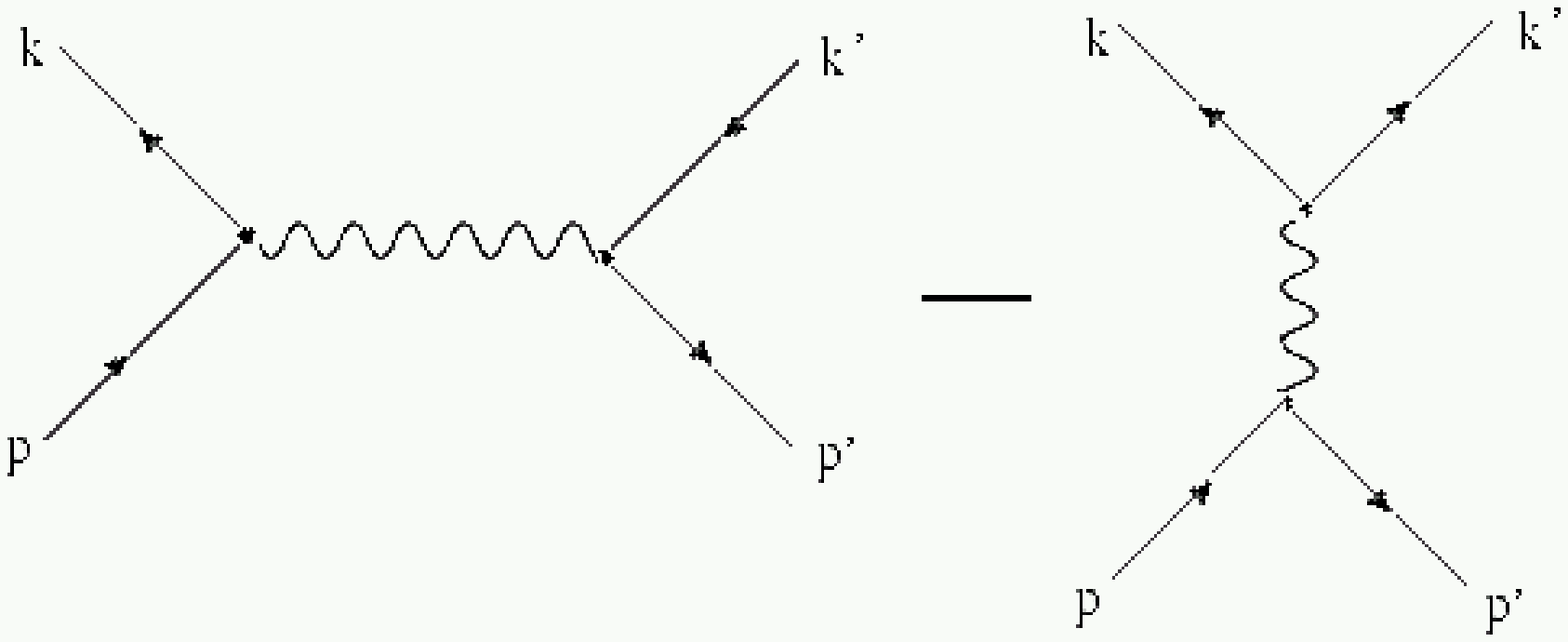}
\caption{\it Tree-level diagrams for Bhabha scattering where
the gauge boson is either a photon or a $Z$.
\label{bha-diag} }
\end{figure}

\begin{figure}[t]
\dofig{3.50in}{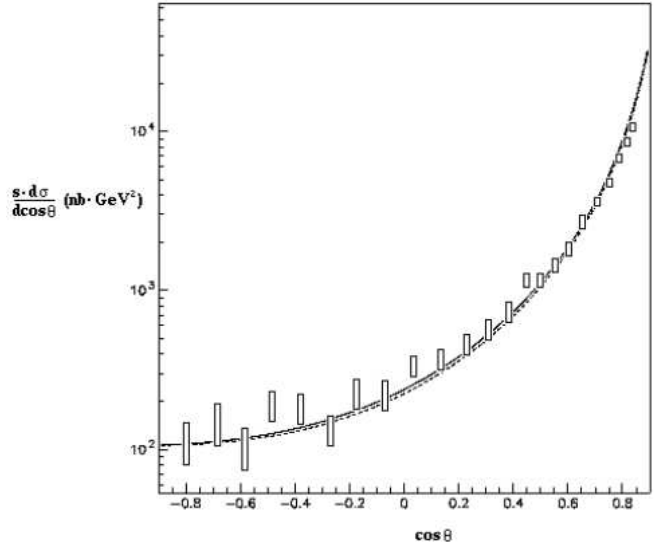}
\caption{\it Bhabha scattering at $\sqrt{s}=9.4~GeV$ as measured by
the PLUTO Collaboration\cite{Berger:1980rq} plotted alongside the SM
 prediction(solid line) and the
NGT prediction (dashed line) for $\epsilon \approx 0.14$. 
\label{bhabha-t-fig} }
\end{figure}

The Bhabha scattering amplitude receives contributions from 
diagrams involving both the photon and the Z boson as shown
in Fig \ref{bha-diag}.
However, it will be simplest to consider this process in the
energy range $m_e \ll \sqrt{s} \ll m_Z$ where we may ignore
the $Z$ diagrams and the electron mass.
In this limit the differential cross section
is approximately (see Appendix for details)
\begin{equation}
\label{bha-cross}
\frac{d\sigma}{d\Omega}_{bSM} + \frac{d\sigma}{d\Omega}_{b1}
+ \frac{d\sigma}{d\Omega}_{b2} 
\end{equation}
where 
\begin{eqnarray}
\frac{d\sigma}{d\Omega}_{bSM} & = & \frac{\alpha^2}{2 s}
\bigg[\frac{1 + \cos^4\theta/2}{\sin^4\theta/2}
+ \frac{1}{2}(1 + \cos^2\theta) - 2 \frac{\cos^4\theta/2}{\sin^2\theta/2}
\bigg]\\
&&\\
\frac{d\sigma}{d\Omega}_{b1} &= &\epsilon \frac{\alpha^2}{ s}
\left[  \frac{\cos^2\theta + 6 \cos\theta - 1}{\sin^2\theta/2}
 +  \sin\theta (\sin\varphi + \cos\varphi)\right]\\
&&\\
\frac{d\sigma}{d\Omega}_{b2} &=& \epsilon^2 \frac{\alpha^2}{2 s}
\bigg[ \sin 2\theta (\sin\varphi + \cos\varphi) - \sin^2 \theta \\
&&\\
&& - 2 \frac{ \cos^2 \frac{\theta}{2}}{\sin^4 \frac{\theta}{2}}
   (\sin \theta (\sin\varphi + \cos\varphi) + \cos \theta + 1) \\
&&\\
&& -4 \frac{ \cos^2 \frac{\theta}{2}}{\sin^2 \frac{\theta}{2}}
 ( \sin\theta \cos\varphi - \cos\theta + 1) \bigg]\\
\end{eqnarray}
Again we see the characteristic oscillatory dependence on the 
azimuthal angle, similar to the prediction from noncommutative
theories\cite{Hewett:2000zp}.
 We can compare the prediction in Eqn \ref{bha-cross}
(setting $\frac{d\sigma}{d\Omega}_{b1}=0$) 
with a measurement by the PLUTO
Collaboration\cite{Berger:1980rq} performed at $\sqrt{s}=9.4~GeV$ (see Figure
\ref{bhabha-t-fig}). We conclude that for
 the case of $\epsilon = 0.14$ there is no
conflict with the data.

\subsection{M$\o$ller Scattering: $e^-e^- \to e^-e^-$}
\label{subsec:moll}
\begin{figure}[t]
\dofig{3.50in}{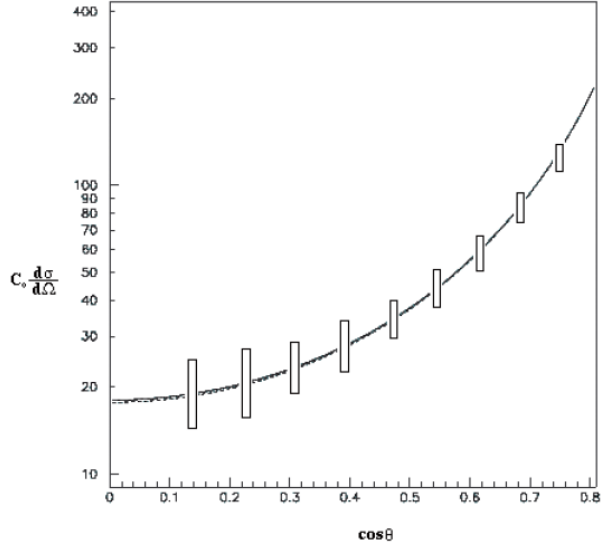}
\caption{\it M$\o$ller scattering at $\sqrt{s}=600~MeV$ at
 Mark III\cite{Barber:1966} plotted on the SM
 prediction(solid line) and the
NGT prediction (dashed line) for $\epsilon \approx 0.14$. The cross
section is plotted in inverse units of 
$C_o \equiv \frac{1}{8}r_D^2 (\frac{m_e}{E})^2$. 
\label{moller-t-fig} }
\end{figure}

Finally we consider the constraints from M$\o$ller scattering.
Now the differential cross section is (see Appendix)
\begin{equation}
\label{moll-cross}
\frac{d\sigma}{d\Omega}_{mSM} + \frac{d\sigma}{d\Omega}_{m1}
+ \frac{d\sigma}{d\Omega}_{m2} 
\end{equation}
where 
\begin{eqnarray}
\frac{d\sigma}{d\Omega}_{mSM} &= & \frac{\alpha^2}{s}
\bigg[ 1 + \frac{1}{\sin^4\theta/2} + \frac{1}{\cos^4\theta/2}
\bigg]\\
&&\\
\frac{d\sigma}{d\Omega}_{m1} &= &-2\epsilon\frac{\alpha^2}{s}
\bigg[  \frac{\cos^2\theta + 6 \cos\theta - 1}{\sin^2\theta/2}
 +  4\frac{\sin^2 \theta/2}{\sin\theta}(\sin\varphi + \cos\varphi)\bigg]\\
&&\\
\frac{d\sigma}{d\Omega}_{m2} &=& -8\epsilon^2 \frac{\alpha^2}
{s \sin^4\theta}
\bigg[ \sin^4 \theta + \sin\theta \cos\theta 
 ((3+ \cos^2 \theta)\sin\varphi + (5 - \cos^2 \theta) \cos\varphi\bigg] \\
\end{eqnarray}
We note a dependence on $\varphi$ similar to that in the
other scattering processes and check the constraint from the
$\theta$ distribution:
Figure \ref{moller-t-fig} shows data taken at the Mark III linear
accelerator at SLAC\cite{Barber:1966}.
Again we see that the agreement between theory and experiment is excellent
for $\epsilon \approx 0.14$.

\section{Constraints from General Relativity}
\label{sec:constraints}
We have seen in the previous sections that NGT
 passes several
 tests in experiments probing high energies. It is relevant
to inquire whether the theory likewise satisfies constraints
at energies corresponding to macroscopic or cosmological
scales. Thus we leave high energy physics to the side for the
moment and concentrate on constraints in the framework of
general relativity. Although a fair number of papers address this
 question we will restrict ourselves to a discussion of only a few
of these. 
\subsection{Newton's $2^{nd}$ Law}
Starting from the geodesic equation in GR,
\begin{equation}
\label{newton}
\frac{d^2 x^\mu}{d \sigma^2} 
+ \Gamma^\mu_{\nu\lambda}\frac{d x^\nu}{d \sigma}
\frac{d x^\lambda}{d \sigma} = 0
\end{equation}
and using the metric chosen in Eqn (\ref{a-form}), we obtain
for a particle in a conservative potential $U(x,y,z)$
the following equations of motion:
\begin{equation}
\label{newton-system}
 m\frac{d^2}{dt^2}
\left(
 \begin{array}{c}
  x \\ y \\ z \\ 
 \end{array}
 \right)
 =
\frac{1}{1 - \epsilon^4}\left(
\begin{array}{ccc}
-1 + \epsilon^2 & -\epsilon(1- \epsilon^2) &  -\epsilon(1+ \epsilon)^2 \\
\epsilon(1- \epsilon^2) & -1+ \epsilon^2 &  -\epsilon(1- \epsilon^2) \\
\epsilon(1- \epsilon)^2 & \epsilon(1- \epsilon^2) & -1+ \epsilon^2 \\ 
\end{array}
\right)
\left(
 \begin{array}{c}
  U_x \\ U_y \\ U_z  \\ 
 \end{array}
 \right)
\end{equation}
where $U_i \equiv dU/dx^i$. Thus motion in a given direction is
influenced by the gradient of the potential in an orthogonal direction,
violating Newton's Second Law\footnotemark.
\footnotetext{Theories of noncommutating coordinates also 
violate Newton's Second Law: 
$\frac{d^2 x^i}{dt^2}= -\frac{dU}{dx^i}+ 
\theta^{ij}\frac{d^2 U}{dx^j dx^k}\frac{d x^k}{dt}$ is the prediction
in these theories\cite{Romero:2002ns}.}
 That planets in the solar system move
on Keplerian orbits to an excellent approximation puts very stringent
constraints on such deviations from $F=ma$. Therefore, if 
$\epsilon$ varies only slowly over solar-system distances, its magnitude
must be vanishingly small to match the observed trajectories of the 
planets. On the other hand, if $\epsilon$ averages to
zero over much smaller  distances($\eg$ sub-micron) then odd powers
of $\epsilon$ may be set to zero in Eqn (\ref{newton-system}). The
even powers of $\epsilon$ may be removed from off-diagonal entries
by a suitable coordinate rotation (see Appendix), reducing
Eqn (\ref{newton-system}) to the usual diagonal form. 
This is the justification for setting odd powers of  $\epsilon$
to zero in the scattering cross section formulae
in Eqns \ref{pair-cross}, \ref{bha-cross}, and \ref{moll-cross}.

\subsection{Cosmology}
Strictly homogeneous and isotropic solutions of the NGT field equations 
always reduce to the Friedmann-Robertson-Walker(FRW) solutions of
GR\cite{Moffat:1997cc}. Relaxing the homogeneous requirement 
slightly ($\ie$ turning on $g_{[\mu\nu]}$) leads to the 
approximate FRW metric
\begin{equation}
ds^2 = dt^2 -R^2(t)
\left(h(r)dr^2 + r^2(d\theta^2 + \sin^2 \theta~d\varphi^2)\right)
\end{equation}
where the field equations determine $h(r)$ as well as $b(r)$
 in the following
equation involving the Hubble variable $H=\dot{R}/R$:
\begin{eqnarray}
H^2(t) + \frac{b(r)}{R^2(t)} &=& \Omega(r,t)H^2(t) \\
 \Omega(r,t) &\equiv& \Omega_M(t) + \Omega_S(r,t) \\
 \Omega_M(t) &=& \frac{8 \pi G \rho_M(t)}{3 H^2(t)}\\
\end{eqnarray}
Here $\Omega_M(t)$ is the usual matter density while $\Omega_S(r,t)$
is the contribution to density from $g_{[\mu\nu]}$. 
Precise measurements of the curvature and density of the universe
can therefore put constraints on the magnitude of $g_{[\mu\nu]}$.

Another way to constrain NGT is by measuring the polarization of
light arriving from distant cosmic sources\cite{Moffat:1997cc}.
This can be seen immediately from the electromagnetic action:
\begin{equation}
I_em \sim \int d^4 x 
\sqrt{-g} g_{\mu\alpha}g_{\nu\beta} F^{\mu\nu}F^{\alpha\beta}
\end{equation}
upon expanding $g$ as in Eqn(\ref{g-components}). One can show 
that terms like $\partial^iA^i\partial^jA^j, i\ne j$ arise
with no counterpart in ordinary electromagnetism. Such terms
imply a ``mixing'' between directions as light propagates,
leading to a distance-dependent polarization.
\subsection{Astrophysics}
Stellar collapse is predicted to differ markedly from the
standard GR prediction: namely, a collapsing star with mass
above the Chandrasekhar limit does not lead to a
 black hole singularity\cite{Moffat:1995pi}. The collapse
is found to asymptotically reach a compact pseduo-stable state which,
like a black hole, emits larges amounts of thermal and gravitational
radiation, but no Hawking radiation. For practical purposes
therefore it may be difficult to distinguish this 
object from a standard GR black hole.

At galactic scales, NGT may provide an explanation for the
flat behaviour of rotation curves in spiral galaxies 
alternative to the conventional theory that the galactic halo consists
of $90$ percent dark matter. NGT can alter Newtonian gravity at
galactic scales\cite{Moffat:1996dq},
 predicting rotation curves in agreement with data, 
without measurably affecting gravity at or below solar system scales. 

We stress that these and other cosmological or astrophysical 
tests of NGT are just as important as ones performed at high
energy such as those considered in this paper. While the latter
tests bear results similar to those from noncommutative theories,
it is the former which can most clearly distinguish between the
two since NGT is a gravitational phenomenon whereas 
noncommutivity in the conventional string theory context is not.

\section{Conclusions}
\label{sec:concl}

We have seen in the preceding analysis that, in electron scattering
 experiments, the predictions of NGT are similar to those from
theories of noncommuting coordinates. Although in both
theories the deviation of the differential cross section  $d\sigma/d\theta$ 
from the SM prediction offers the strongest constraint, we suggest
that  as experimental precision improves
 the oscillatory behaviour of 
 $d\sigma/d\varphi$ should be the clearest prediction of these theories
since the SM background is flat.
We conclude that the OPAL data is consistent with both
a ncSM with $\theta_{\mu\nu} \sim (141~ GeV)^{-2}$ and
the particular NGT presented in this paper for values
of $\epsilon \sim~0.14$, implying that the metric of space-time
could be up to $2$ percent antisymmetric in the neighborhood
of terrestrial experiments. Precision data from PLUTO
and MarkIII confirm the latter bound on NGT. 
However, in contrast to the parameter $\epsilon$,
 which is dimensionless,
the  parameter  $\theta_{\mu\nu}$ is of mass dimension $-2$ 
and therefore should
cause deviations from the SM which 
scale with the square of center-of-mass energy.
 As more data 
becomes available from high energy collider experiments such
as those planned at the LHC or a future $e^+ e^-$ linear collider,
noncommutative signals should therefore grow stronger and eventually
overtake those of NGT.

From a theoretical perspective, NGT is a much cleaner theory than
the ncSM. The latter, aside from some difficulties in the gauge 
sector, suffers from severe Lorentz violation and a still unsolved
problem in ultraviolet divergences\cite{Hinchliffe:2002km}.
Nonetheless, both theories represent interesting perturbations 
of ordinary space-time. Mixing of space-time coordinates is
a common feature of both noncommutative theory and NGT,
though
the origin of this mixing arises from electromagnetism in the
former and gravity in the latter. Cosmological and astrophysical 
phenomenology should therefore readily distinguish the two theories.
Work in this area has thus  far been encouraging.

\section*{Acknowledgements}
 This work was supported by the Department of Physics at Tsinghua University.

\section*{Appendix}
\subsection*{Useful Identities}
The following is a partial list of contractions and 
trace identities for gamma matrices contracted with $a_{\mu\nu}$.
\subsubsection*{contraction identities}
\begin{eqnarray}
\gamma_\mu &=& (\eta_{\mu\nu} + a_{\mu\nu})\gamma^\nu \\
\gamma^\mu\gamma_\mu &=& 4 + \gamma^\mu a_{\mu\nu}\gamma^\nu \\
\gamma_\mu\gamma^\mu  &=& 4 -  \gamma^\mu a_{\mu\nu}\gamma^\nu \\
\gamma^\mu \gamma^\nu \gamma_\mu &=&  -2 \gamma^\nu
 - \gamma^\nu  \gamma^\mu a_{\mu\rho}\gamma^\rho
   + 2 \eta^{\nu\mu} a_{\mu\rho}\gamma^\rho \\
 \gamma_\mu\gamma^\nu\gamma^\mu &=&  -2 \gamma^\nu
 +  \gamma^\mu a_{\mu\rho}\gamma^\rho \gamma^\nu
 + 2 \eta^{\nu\mu} a_{\mu\rho}\gamma^\rho \\
\gamma^\mu \gamma^\nu  \gamma^\rho \gamma_\mu &=& 4 \eta^{\nu\rho}
+  \gamma^\nu  \gamma^\rho  \gamma^\mu a_{\mu\lambda}\gamma^\lambda
+ 2(\gamma^\rho \eta^{\nu\mu} -
\gamma^\nu\eta^{\rho\mu}) a_{\mu\lambda}\gamma^\lambda \\
 \gamma_\mu \gamma^\nu \gamma^\rho \gamma^\mu  &=&
4 \eta^{\nu\rho} - 
  \gamma^\mu a_{\mu\lambda}\gamma^\lambda \gamma^\nu  \gamma^\rho
 +2( \eta^{\rho\mu} a_{\mu\lambda} \gamma^\lambda \gamma^\nu -
  \eta^{\nu\mu} a_{\mu\lambda} \gamma^\lambda \gamma^\rho)\\ 
\gamma^\mu \gamma^\nu  \gamma^\rho  \gamma^\sigma \gamma_\mu &=&
-2  \gamma^\sigma  \gamma^\rho \gamma^\nu - 
\gamma^\nu  \gamma^\rho  \gamma^\sigma 
\gamma^\mu a_{\mu\lambda}\gamma^\lambda
+ 2( \gamma^\nu\gamma^\rho \eta^{\sigma\mu} -  
      \gamma^\nu\gamma^\sigma\eta^{\rho\mu}
   + \gamma^\rho\gamma^\sigma  
      \eta^{\nu\mu})a_{\mu\lambda}\gamma^\lambda 
\nonumber\\
\end{eqnarray}
\subsubsection*{trace identities}
\begin{eqnarray}
tr[\gamma^\mu a_{\mu\nu}\gamma^\nu] &=& 0 \\
tr[\gamma^\lambda a_{\lambda\mu}\gamma^\nu] &=&
  4 \eta^{\nu\rho}a_{\rho\mu}\\ 
tr[\gamma^\mu a_{\mu\nu}\gamma^\nu\gamma^\sigma \gamma^\tau] &=&
-8 \eta^{\sigma\mu}a_{\mu\nu} \eta^{\nu\tau} \\
tr[\gamma^\mu a_{\mu\nu}\gamma^\nu
\gamma^\rho\gamma^\sigma \gamma^\tau\gamma^\lambda] &=&
8( -\eta^{\rho\mu}a_{\mu\nu} \eta^{\nu\sigma}\eta^{\tau\lambda}
+ \eta^{\rho\mu}a_{\mu\nu} \eta^{\nu\tau}\eta^{\sigma\lambda}
- \eta^{\rho\mu}a_{\mu\nu} \eta^{\nu\lambda}\eta^{\sigma\tau}
\nonumber\\
&&
-  \eta^{\sigma\mu}a_{\mu\nu} \eta^{\nu\tau} \eta^{\rho\lambda}
+ \eta^{\sigma\mu}a_{\mu\nu} \eta^{\nu\lambda} \eta^{\rho\tau}
- \eta^{\tau\mu}a_{\mu\nu} \eta^{\nu\lambda} \eta^{\rho\sigma}
\nonumber\\
\end{eqnarray}

\subsection*{Pair Annihilation}
To compute the pair annihilation cross section, we may
start from the expression for Compton scattering which is related
by crossing symmetry ($p_1 \to p, p_2 \to -p^\prime, k_1 \to -k, 
k_2 \to k^\prime$):
\begin{eqnarray}
i{\cal
M}&=&\bar{u}(p^\prime)(-ie\gamma^\mu)\epsilon^\ast_\mu(k^\prime)\frac{i(p\hspace{-0.17cm}\slash+k\hspace{-0.17cm}\slash+m)}{(p+k)^2-m^2}(-ie\gamma^\nu)\epsilon_\nu(k)u(p)\nonumber\\
&&+\bar{u}(p^\prime)(-ie\gamma^\nu)\epsilon_\nu(k)\frac{i(p\hspace{-0.17cm}\slash-k\hspace{-0.17cm}\slash^\prime+m)}{(p-k^\prime)^2-m^2}(-ie\gamma^\mu)\epsilon^\ast_\mu(k^\prime)u(p)\nonumber\\
&=&-ie^2\epsilon^\ast_\mu(k^\prime)\epsilon_\nu(k)\bar{u}(p^\prime)\bigg[\frac{\gamma^\mu(p\hspace{-0.17cm}\slash+k\hspace{-0.17cm}\slash+m)\gamma^\nu}{(p+k)^2-m^2}+\frac{\gamma^\nu(p\hspace{-0.17cm}\slash-k\hspace{-0.17cm}\slash^\prime+m)\gamma^\mu}{(p-k^\prime)^2-m^2}\bigg]u(p)
\end{eqnarray}
From here,
the cross section is
\begin{eqnarray}
{1\over4}\sum_{spins}|{\cal
M}|^2&=&{e^4\over4}g_{\mu\rho}g_{\nu\sigma}\cdot
tr\bigg\{(p\hspace{-0.17cm}\slash^\prime+m)\bigg[\frac{\gamma^\mu
k\hspace{-0.17cm}\slash\gamma^\nu+2\gamma^\mu p^\nu}{2p\cdot
k}+\frac{\gamma^\nu
k\hspace{-0.17cm}\slash^\prime\gamma^\mu-2\gamma^\nu
p^\mu}{2p\cdot k^\prime}\bigg]\nonumber\\
&&\cdot(p\hspace{-0.17cm}\slash+m)\bigg[\frac{\gamma^\sigma
k\hspace{-0.17cm}\slash\gamma^\rho+2\gamma^\rho p^\sigma}{2p\cdot
k}+\frac{\gamma^\rho
k\hspace{-0.17cm}\slash^\prime\gamma^\sigma-2\gamma^\sigma
p^\rho}{2p\cdot k^\prime}\bigg]\bigg\}\nonumber\\
&\equiv&{e^4\over4}\bigg[\frac{I}{(2p\cdot
k)^2}+\frac{II}{(2p\cdot k)(2p\cdot k^\prime)}+\frac{III}{(2p\cdot
k^\prime )(2p\cdot k)}+\frac{IV}{(2p\cdot k^\prime)^2}\bigg]
\end{eqnarray}
 where the first trace is
\begin{eqnarray}
I&=&tr[(p\hspace{-0.17cm}\slash^\prime+m)(\gamma^\mu
k\hspace{-0.17cm}\slash\gamma^\nu+2\gamma^\mu
p^\nu)(p\hspace{-0.17cm}\slash+m)(\gamma_\nu
k\hspace{-0.17cm}\slash\gamma_\mu+2\gamma_\mu p_\nu)]\nonumber\\
\end{eqnarray}
First we consider the ${\cal O}(\epsilon)$ contributions to this
amplitude.
In the computation we will need  to insert the full expression
 for $g_{\mu\nu}$ only; note that to leading order Dirac 
propagators are unchanged,
 $\eg$ $\Sigma_s u(p)\overline{u}_s(p)=
 p\hspace{-0.17cm}\slash+m \equiv p^{\mu}\eta_{\mu\nu}\gamma^{\nu}+m$,
 which is true even after radiative corrections.
After much Dirac algebra using the contraction and trace identities
above and taking the high energy limit $m=0$,
we find that $I$ has no ${\cal O}(\epsilon)$ piece.
However the second trace,
\begin{eqnarray}
II&=&tr[(p\hspace{-0.17cm}\slash^\prime+m)(\gamma^\mu
k\hspace{-0.17cm}\slash\gamma^\nu+2\gamma^\mu
p^\nu)(p\hspace{-0.17cm}\slash+m)(\gamma_\mu
k\hspace{-0.17cm}\slash^\prime\gamma_\nu-2\gamma_\nu p_\mu)]\nonumber\\
&=&tr[p\hspace{-0.17cm}\slash^\prime\gamma^\mu
k\hspace{-0.17cm}\slash\gamma^\nu
p\hspace{-0.17cm}\slash\gamma_\mu
k\hspace{-0.17cm}\slash^\prime\gamma_\nu]-2tr[p\hspace{-0.17cm}\slash^\prime\gamma^\mu
k\hspace{-0.17cm}\slash\gamma^\nu p\hspace{-0.17cm}\slash\gamma_\nu p_\mu]\nonumber\\
&&+mtr[p\hspace{-0.17cm}\slash^\prime\gamma^\mu
k\hspace{-0.17cm}\slash\gamma^\nu\gamma_\mu
k\hspace{-0.17cm}\slash^\prime\gamma_\nu]-2mtr[p\hspace{-0.17cm}\slash^\prime\gamma^\mu
k\hspace{-0.17cm}\slash\gamma^\nu\gamma_\nu p_\mu]\nonumber\\
&&+2tr[p\hspace{-0.17cm}\slash^\prime\gamma^\mu p^\nu
p\hspace{-0.17cm}\slash\gamma_\mu
k\hspace{-0.17cm}\slash^\prime\gamma_\nu]-4tr[p\hspace{-0.17cm}\slash^\prime\gamma^\mu
p^\nu p\hspace{-0.17cm}\slash\gamma_\nu
p_\mu]\nonumber\\
&&+2mtr[p\hspace{-0.17cm}\slash^\prime\gamma^\mu p^\nu\gamma_\mu
k\hspace{-0.17cm}\slash^\prime\gamma_\nu]-4mtr[p\hspace{-0.17cm}\slash^\prime\gamma^\mu p^\nu\gamma_\nu p_\mu]\nonumber\\
&&+mtr[\gamma^\mu
k^\prime\hspace{-0.17cm}\slash\gamma^\nu\gamma_\mu
k\hspace{-0.17cm}\slash^\prime\gamma_\nu]-2mtr[\gamma^\mu
k\hspace{-0.17cm}\slash\gamma^\nu p\hspace{-0.17cm}\slash\gamma_\nu p_\mu]\nonumber\\
&&+m^2tr[\gamma^\mu k\hspace{-0.17cm}\slash\gamma^\nu\gamma_\mu
k\hspace{-0.17cm}\slash^\prime\gamma_\nu]-2m^2tr[\gamma^\mu
k\hspace{-0.17cm}\slash\gamma^\nu\gamma_\nu
p_\mu]\nonumber\\
&&+2mtr[\gamma^\mu p^\nu p\hspace{-0.17cm}\slash\gamma_\mu
k\hspace{-0.17cm}\slash^\prime\gamma_\nu]-4mtr[\gamma^\mu p^\nu
p\hspace{-0.17cm}\slash\gamma_\nu p_\mu]\nonumber\\
&&+2m^2tr[\gamma^\mu p^\nu\gamma_\mu
k\hspace{-0.17cm}\slash^\prime\gamma_\nu]-4m^2tr[\gamma^\mu p^\nu\gamma_\nu p_\mu]\nonumber\\
&=&tr[p\hspace{-0.17cm}\slash^\prime\gamma^\mu
k\hspace{-0.17cm}\slash\gamma^\nu
p\hspace{-0.17cm}\slash\gamma_\mu
k\hspace{-0.17cm}\slash^\prime\gamma_\nu]-2tr[p\hspace{-0.17cm}\slash^\prime\gamma^\mu
k\hspace{-0.17cm}\slash\gamma^\nu p\hspace{-0.17cm}\slash\gamma_\nu p_\mu]\nonumber\\
&&+2tr[p\hspace{-0.17cm}\slash^\prime\gamma^\mu p^\nu
p\hspace{-0.17cm}\slash\gamma_\mu
k\hspace{-0.17cm}\slash^\prime\gamma_\nu]-4tr[p\hspace{-0.17cm}\slash^\prime\gamma^\mu
p^\nu p\hspace{-0.17cm}\slash\gamma_\nu
p_\mu]\nonumber\\
\end{eqnarray}
yields
\begin{eqnarray}
II&=&\mbox{ordinary theory}\nonumber\\
&&-32\epsilon p^\prime\cdot pk\oslash k^\prime+16\epsilon(p\cdot
kp\oslash p^\prime +p^\prime\cdot pk\oslash p)
\end{eqnarray}
where we define
\begin{eqnarray}
p\cdot k &\equiv& p_\mu \eta^{\mu\nu} k_\nu \nonumber \\
p\oslash k &\equiv&
 p_\mu \eta^{\mu\nu} a_{\nu\rho} \eta^{\rho\sigma} k_\sigma \nonumber \\ 
\end{eqnarray}
and ``ordinary theory'' refers to the SM ($\epsilon=0$).
The other traces are trivially obtained from the above:
\begin{eqnarray}
&&III=II\nonumber\\
&&IV=I(k\rightarrow-k^\prime)
\end{eqnarray}
Transforming to the pair annihiliation momenta by crossing
symmetry,
\begin{eqnarray}
p\rightarrow
p_1,p^\prime\rightarrow-p_2,k\rightarrow-k_1,k^\prime\rightarrow
k_2
\end{eqnarray}
we obtain
\begin{eqnarray}
{1\over4}\sum_{spins}|{\cal
M}|^2&\equiv&{e^4\over4}\bigg[\frac{I}{(2p_1\cdot
k_1)^2}-\frac{II}{(2p_1\cdot k_1)(2p_1\cdot
k_2)}-\frac{III}{(2p_1\cdot k_2 )(2p_1\cdot
k_1)}+\frac{IV}{(2p_1\cdot k_2)^2}\bigg]\nonumber\\
\end{eqnarray}
and
\begin{eqnarray}
II&=&III\nonumber\\
&=&\mbox{ordinary theory}\nonumber\\
&&-32\epsilon p_2\cdot p_1k_1\oslash k_2+16\epsilon(p_1\cdot
k_1p_1\oslash p_2 +p_2\cdot p_1k_1\oslash p_1)
\end{eqnarray}
\begin{eqnarray}
I&=&\mbox{ordinary theory}
\end{eqnarray}
\begin{eqnarray}
IV&=&\mbox{ordinary theory}
\end{eqnarray}
Define the kinematical variables $E, \theta, \varphi$ in the
center of mass frame:
\begin{eqnarray}
\label{kin-vars}
&&p_1=(E,E\hat{z})\nonumber\\
&&p_2=(E,-E\hat{z})\nonumber\\
&&k_1=(E,E\sin\theta\cos\varphi,E\sin\theta\sin\varphi,E\cos\theta)\nonumber\\
&&k_2=(E,-E\sin\theta\cos\varphi,-E\sin\theta\sin\varphi,-E\cos\theta)
\end{eqnarray}
It is now straightforward to derive the ${\cal O}(\epsilon)$
differential cross section:
\begin{eqnarray}
\frac{d
\sigma}{d\Omega}&=&\mbox{ordinary
theory}\nonumber\\
&&+\frac{\alpha^2\epsilon}{2E^2}\cdot\bigg[\frac{2\cos\theta}{\sin^2\theta}+\frac{(\sin\varphi+\cos\varphi)}{\sin\theta}\bigg]\nonumber\\
\end{eqnarray}
Now for  ${\cal O}(\epsilon^2)$:
The first trace eventually reduces to
\begin{eqnarray}  
I&=&32[p\cdot k p^\prime\otimes k+p^\prime\cdot kp\otimes
k-p^\prime\cdot pk\otimes k]\nonumber\\
&&-32p^\prime\oslash kp\oslash k -16tr(a^2)p^\prime\cdot kp\cdot k
\end{eqnarray}
where we introduce the notation $\otimes$:
\begin{eqnarray}
k\otimes p&=&k_\mu
\eta^{\mu\nu} a_{\nu\rho}\eta^{\rho\sigma}a_{\sigma\tau}
\eta^{\tau\lambda}p_\lambda =p\otimes k \nonumber\\
tr(a^2) &=& \eta^{\mu\nu} a_{\nu\rho}
  \eta^{\rho\sigma}a_{\sigma\tau}\eta^\tau_{\cdot\mu} \nonumber\\
\end{eqnarray}
Similarly, the second trace is
\begin{eqnarray}
II&=&16\epsilon^2(k^\prime\otimes kp^\prime\cdot p+p^\prime\otimes
pk^\prime\cdot
k )\nonumber\\
\end{eqnarray}
Then the other traces follow easily as before:
\begin{eqnarray}
III&=&II=16\epsilon^2(k^\prime\otimes kp^\prime\cdot
p+p^\prime\otimes pk^\prime\cdot k )
\end{eqnarray}
\begin{eqnarray}
IV&=&I(k\rightarrow-k^\prime)=32[p\cdot k^\prime p^\prime\otimes
k^\prime+p^\prime\cdot k^\prime p\otimes
k^\prime-p^\prime\cdot pk^\prime\otimes k^\prime]\nonumber\\
&&-32p^\prime\oslash k^\prime p\oslash k^\prime
-16tr(a^2)p^\prime\cdot k^\prime p\cdot k^\prime
\end{eqnarray}
From the particular choice of $a$ in Eqn \ref{a-form}, we have
\begin{eqnarray}
 a_{\nu\rho}\eta^{\rho\sigma}a_{\sigma\tau}
&=&
\left(
\begin{array}{cccc}
3&2&0&-2\\
2&1&0&-2\\
0&0&1&0\\
-2&-2&0&1
\end{array}
\right)
\end{eqnarray}
whence it follows from the kinematical assignments that
\begin{eqnarray}
{1\over4}\sum_{spins}|{\cal
M}|^2&=&{\rm ordinary} + {\cal O}(\epsilon) + 
4e^4\epsilon^2\bigg[\frac{\cos^2\theta-3}{\sin^2\theta}+\frac{(1+\cos\theta)\sin2\theta}{\sin^4\theta}(\sin\varphi-\cos\varphi)-\frac{\sin2\theta}{\sin^2\theta}\cos\varphi\bigg]\nonumber\\
\end{eqnarray}
leading to the expression reported in Eqn \ref{pair-cross}.

\subsection*{Bhabha Scattering}
The two diagrams contributing (see Figure \ref{bha-diag})
 interfere destructively,
with spin averaged squared amplitude
\begin{eqnarray}
{1\over4}\sum_{spins}|{\cal
M}|^2&=&
\frac{e^4}{4}\bigg[\frac{I}{(p+p^\prime)^4}
 + \frac{II}{(k^\prime-p^\prime)^4} 
- \frac{III + IV}{(p+p^\prime)^2 (k^\prime-p^\prime)^2}
\bigg] \nonumber\\
I &\equiv& tr[p\hspace{-0.17cm}\slash^\prime \gamma^\mu
 p\hspace{-0.17cm}\slash \gamma^\sigma]
tr[  k\hspace{-0.17cm}\slash \gamma^\nu 
k\hspace{-0.17cm}\slash^\prime\gamma^\rho]g_{\mu\nu}g_{\rho\sigma}
\nonumber\\
II &\equiv& I(p \leftrightarrow k^\prime) \nonumber\\
III  &\equiv& tr[p\hspace{-0.17cm}\slash^\prime \gamma^\mu
 p\hspace{-0.17cm}\slash  \gamma^\rho 
 k\hspace{-0.17cm}\slash  \gamma^\nu
k\hspace{-0.17cm}\slash^\prime \gamma^\sigma]g_{\mu\nu}g_{\rho\sigma}
 \nonumber\\
IV &\equiv& III(p \leftrightarrow k^\prime) \nonumber\\
\end{eqnarray}
Then
\begin{eqnarray}
I &=&  p^\prime_\alpha p_\beta (\eta^{\alpha \mu}\eta^{\beta \sigma}
 - \eta^{\alpha \beta}\eta^{\mu \sigma}+
\eta^{\alpha \sigma}\eta^{\mu\beta})\oslash
k_\lambda k^\prime_\tau
(\eta^{\lambda \nu}\eta^{\tau \rho} -
\eta^{\lambda \tau}\eta^{\nu \rho} +
\eta^{\lambda \rho}\eta^{\nu \tau})g_{\mu\nu}g_{\rho\sigma}\nonumber\\
&=& {\rm ordinary theory} + \nonumber\\
&& -2 \bigg[(p \oslash k^\prime)(p^\prime \oslash k)
+ (p^\prime \oslash k^\prime)(p \oslash k) +
(k \cdot k^\prime)(p \otimes p^\prime)\bigg] \nonumber\\
\end{eqnarray}
($\ie$ there is no ${\cal O}(\epsilon)$ piece).
By repeated use of the contraction identities noted earlier, one
can further verify that
\begin{eqnarray}
III+IV &=& {\rm ordinary~theory} + 
32 \bigg[ (p \cdot p^\prime)(k \oslash k^\prime) +
 (p^\prime \cdot k^\prime)(k \oslash p)
+  (p \cdot k^\prime)  (p^\prime \otimes k) -
   (k \cdot p^\prime)  (k^\prime \otimes p)\bigg]\nonumber\\
\end{eqnarray}
Using the kinematical assignments as in Eqn \ref{kin-vars},
replacing $p_1 \to p, p_2 \to p^\prime, k_1 \to k, k_2 \to k^\prime$,
the quoted cross section in the text follows straightforwardly.

\subsection*{M$\o$ller Scattering}
This can be obtained quickly from Bhabha scattering by 
substituting $p$ for $k$ and vice versa in the traces ( but not in
the denominators). Hence
\begin{eqnarray}
{1\over4}\sum_{spins}|{\cal
M}|^2&=&
\frac{e^4}{4}\bigg[\frac{I}{(p+p^\prime)^4}
 + \frac{II}{(k^\prime-p^\prime)^4} 
- \frac{III + IV}{(p+p^\prime)^2 (k^\prime-p^\prime)^2}
\bigg] \nonumber\\
\end{eqnarray}
where
\begin{eqnarray}
I &=&  {\rm ordinary~theory} + \nonumber\\
&& -2 \bigg[(k \oslash k^\prime)(p^\prime \oslash p)
+ (p^\prime \oslash k^\prime)(k \oslash p) +
(p \cdot k^\prime)(k \otimes p^\prime)\bigg] \nonumber\\
II &=& I(k \leftrightarrow k^\prime) \nonumber\\
III+IV &=&  {\rm ordinary~theory} + 
32 \bigg[ (k \cdot p^\prime)(p \oslash k^\prime) +
 (p^\prime \cdot k^\prime)(k \oslash p)
+  (k \cdot k^\prime)  (p^\prime \otimes p) -
   (p \cdot p^\prime)  (k^\prime \otimes k)\bigg]\nonumber\\
\end{eqnarray}
Making the same kinematical assignments as for Bhabha scattering,
the quoted cross section in Section \ref{subsec:moll} follows.

\subsection*{Preserving $F=ma$}
Starting from Eqn(\ref{newton-system}), perform a rotation in
the $x-z$ plane:
\begin{equation}
\label{rotated}
 \frac{d^2}{dt^2}
\left(
 \begin{array}{c}
  x' \\ y' \\ z' \\ 
 \end{array}
 \right)
 =
{\bf \Omega}
\left(
 \begin{array}{c}
  U_x' \\ U_y' \\ U_z'  \\ 
 \end{array}
 \right)
\end{equation}
where
\begin{equation}
{\bf \Omega} = 
\frac{1}{1 - \epsilon^4}\left(
\begin{array}{ccc}
\epsilon^2 - 1 - 4\epsilon^2 c_\theta s_\theta &
\frac{1-\epsilon^4 }{1+\epsilon^2}\epsilon (s_\theta- c_\theta) &
\epsilon( 1 + 2\epsilon(c^2_\theta - s^2_\theta) + \epsilon^2) \\
-\frac{1-\epsilon^4 }{1+\epsilon^2}\epsilon (s_\theta- c_\theta) &
-\frac{1-\epsilon^4 }{1+\epsilon^2} &
-\frac{1-\epsilon^4 }{1+\epsilon^2}\epsilon (s_\theta+ c_\theta) \\
-\epsilon( 1 + 2\epsilon(s^2_\theta - c^2_\theta) + \epsilon^2) &
\epsilon\frac{1-\epsilon^4 }{1+\epsilon^2} &
\epsilon^2 - 1 + 4\epsilon^2 c_\theta s_\theta \\
\end{array}
\right)
\end{equation}
Neglecting odd powers of $\epsilon$, this can clearly be brought
to diagonal form by setting $ c_\theta =  s_\theta$,
 $\ie$ rotating by 45 degrees.

\bibliography{all.bib}
\bibliographystyle{unsrt}

\end{document}